\documentclass[aps,prb,twocolumn,showpacs]{revtex4}
\usepackage{graphicx}

\begin{document}
\title{Doping dependence of the gap anisotropy in
La$_{2-x}$Ce$_{x}$CuO$_{4}$ \\ studied by millimeter-wave
spectroscopy}
\author{A. V. Pronin,$^{1,2,}\footnote {Electronic address:
pronin@ran.gpi.ru, artem.pronin@physik.uni-augsburg.de}$ A.
Pimenov,$^{1}$ A. Loidl,$^{1}$ A. Tsukada,$^{3}$ and M.
Naito$^{3}$}
\address{$^{1}$Experimentalphysik V, Universit\"{a}t Augsburg, 86135 Augsburg,
Germany \\ $^{2}$Institute of General Physics, Russian Academy of
Sciences, 119991 Moscow, Russia \\ $^{3}$NTT Basic Research
Laboratories, 3-1 Morinosato Wakamiya, Atsugi-shi, Kaganawa 243,
Japan}

Submitted to PRB: December 26, 2002

\begin{abstract}
We measure the penetration depth of optimally doped and underdoped
La$_{2-x}$Ce$_{x}$CuO$_{4}$ in the millimeter frequency domain (4
- 7 cm$^{-1}$) and for temperatures 2 K $\leq T \leq$ 300 K. The
penetration depth as function of temperature reveals significant
changes on electron doping. It shows quadratic temperature
dependence in underdoped samples, but increases almost
exponentially at optimal doping. Significant changes in the gap
anisotropy (or even in the gap symmetry) may account for this
transition.
\end{abstract}


\maketitle

Being one of the keystones for theories of high-temperature
superconductivity, the symmetry and the actual shape of the gap
function in high-$T_{c}$'s is one of most attractive issues for
experimental studies. At present the $d$-wave pairing symmetry in
the hole doped materials seems to be well
established.\cite{annett1, tsuei1} But so far there is no common
consensus on the electron-doped cuprates. It was initially
suggested that these materials reveal an $s$-wave
symmetry,\cite{wu, anlage, alff, kashiwaya} however, later
measurements supported strongly a $d$-wave scenario.\cite{tsuei2,
kokales, prozorov, armitage, sato}

A number of recent reports indicate possible transition in the gap
symmetry with doping both in hole \cite{deutscher, yeh, sharoni}
and electron \cite{skinta, biswas} doped cuprates. Tunnelling data
by Dagan and Deutscher, \cite{deutscher} and by Sharoni {\it et
al.} \cite{sharoni} suggest a possible transition from pure
$d_{x^{2}-y^{2}}$ to $d_{x^{2}-y^{2}}+id_{xy}$ or
$d_{x^{2}-y^{2}}+is$ symmetry while changing from the under- to
the overdoped regime in YBa$_{2}$Cu$_{3}$O$_{7-\delta}$, as well
as in Y$_{1-x}$Ca$_{x}$Ba$_{2}$Cu$_{3}$O$_{7-\delta}$. This
transition could be due to surface effects, but possibly reflects
a bulk property, namely the existence of a quantum critical point
in these materials.\cite{laughlin, vojta, khveshchenko} The paper
by Yeh {\it et al.}\cite{yeh} reports on changing of the paring
symmetry from predominately $d_{x^{2}-y^{2}}$ in the under- and
optimally doped YBa$_{2}$Cu$_{3}$O$_{7-\delta}$ single crystals to
$d_{x^{2}-y^{2}}+s$ in the overdoped
Y$_{1-x}$Ca$_{x}$Ba$_{2}$Cu$_{3}$O$_{7-\delta}$ thin films. That
is, there is no apparent breaking of the time-reversal symmetry.

The point contact spectroscopy data on electron doped
Pr$_{2-x}$Ce$_{x}$CuO$_{4}$ by Biswas {\it et al.} \cite{biswas}
also indicate different pairing symmetries in samples with
different doping: $d$-wave in the underdoped and $s$-wave in the
overdoped case. The radio-frequency inductance measurements by
Skinta {\it et al.} \cite{skinta} show a tiny, but measurable
difference in the low temperature ($<$ 3 K) behavior of the
penetration depth between underdoped, optimally doped, and
overdoped samples of Pr$_{2-x}$Ce$_{x}$CuO$_{4}$ and
La$_{2-x}$Ce$_{x}$CuO$_{4}$, indicating $d$-wave symmetry in the
underdoped, and $s$-wave symmetry in the optimally doped and
overdoped regimes.

Overall the majority of these studies is quite consistent: the
underdoped samples tend to demonstrate pure $d$-wave symmetry,
while the overdoped samples rather reveal $d+id_{xy}$, $d+is$ or
$s$-wave-like behavior. Optimal doping stays at the border of
these two possibilities, falling more into the pure $d$-wave limit
in the hole-doped compounds, \cite{yeh} and into the conventional
$s$-wave-like behavior in the electron-doped ones. \cite{skinta}

Here we present measurements of the superconducting penetration
depth $\lambda$ of optimally doped and underdoped electron
superconductors La$_{2-x}$Ce$_{x}$CuO$_{4}$ by a quasi-optical
method at millimeter wavelengths. \cite{kozlov} We show that the
temperature dependence of $\lambda$ sensitively depends on the
doping concentration.

High quality La$_{2-x}$Ce$_{x}$CuO$_{4}$ films with $x=0.106$
(optimally doped) and $x=0.081$ (underdoped), have been deposited
by molecular-beam epitaxy \cite{naito1, naito2} on transparent
(001)-oriented SrLaAlO$_{4}$ substrates. The substrates were
plane-parallel plates, approximately $10\times10$ mm in size with
thickness of 0.335 mm for underdoped and of 0.472 mm for optimally
doped films. The film thickness was 140 nm. The films had sharp
resistive and inductive transitions at $T_c = 30$\,K (optimally
doped sample) and 25 K (underdoped sample) with $\Delta T_{c} <
1$\,K.

The requirements to prepare high-quality
La$_{2-x}$Ce$_{x}$CuO$_{4}$ films are stringent cation
stoichiometry adjustment and careful removal of the apical oxygen
without phase decomposition. With regard to the latter issue, we
kept films at 600 $^{\circ}$C for 10 min in the oxygen pressure
less than 10$^{-8}$ Torr just after the film growth.  By this
reduction recipe, the superconducting properties were optimized.
No phase decomposition was detected in the reflection high-energy
electron diffraction (RHEED) measurements. Although it is
generally very difficult to evaluate the exact oxygen content in
films, our {\it in-situ} Cu $2p$ x-ray photoemission spectra on
the film surfaces indicate that the Cu ions are mostly coordinated
by four O ions in a square planar configuration (no apical
oxygen). Similar studies are reported for
Nd$_{2-x}$Ce$_{x}$CuO$_{4}$ films in Ref. \onlinecite{yamamoto}.

\begin{figure}[t]
\centering
\includegraphics[width=\columnwidth,clip]{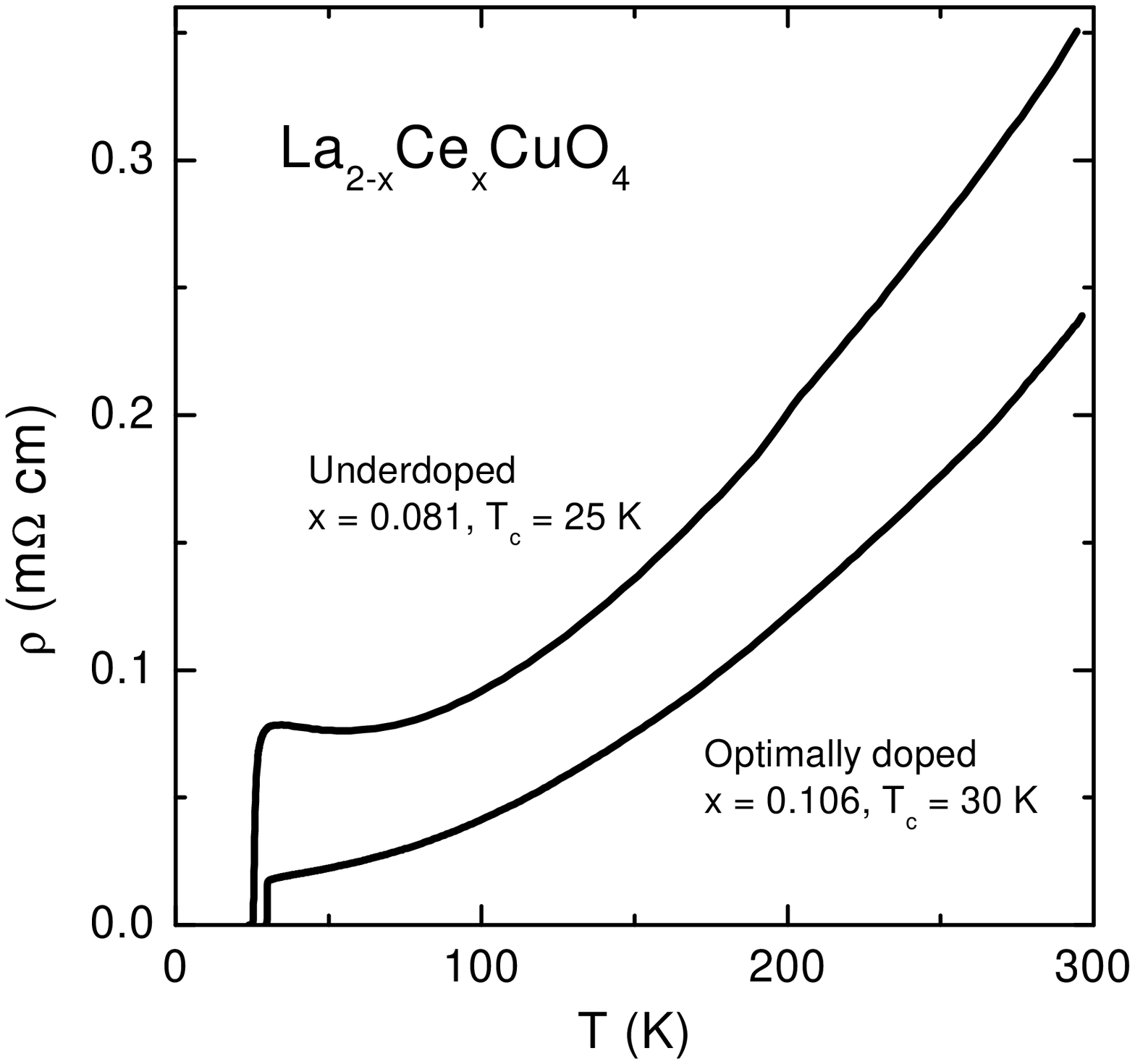}
\caption{Temperature dependence of dc resistivity for the
optimally doped (bottom line) and the underdoped (upper line)
La$_{2-x}$Ce$_{x}$CuO$_{4}$ samples.}
\label{resistivity}
\end{figure}

The temperature dependence of dc resistivity of both films, as
measured by the standard four probe method with evaporated Ag
electrodes, is shown in  Fig. \ref{resistivity}. The overall shape
of the $\rho(T)$ curves is quadratic rather than linear. The
underdoped sample reveals a slight upturn in $\rho(T)$ at low
temperatures ($\lesssim 70$\,K). This is an indication of
localization effects, and is quite typical for this compound in
the underdoped regime. \cite{naito2}

The measurements have been performed with a quasi-optical coherent
source (backward-wave oscillator) spectrometer, \cite{kozlov}
working in the millimeter to submillimeter wavelength domain. The
Mach-Zehnder interferometer arrangement allows measuring both, the
intensity and the phase shift of the wave transmitted through the
La$_{2-x}$Ce$_{x}$CuO$_{4}$ films on the substrates. Using the
Fresnel optical formulas for the complex transmission coefficient
of the substrate-film system, both components of the in-plane
dynamic conductivity ($\sigma_{1} + i\sigma_{2}$) of the films
have been directly calculated from these measurements. The
penetration depth has been determined as
$\lambda={c/(4\pi\omega\sigma_{2})^{1/2}}$, where $c$ is the light
velocity and $\omega$ the angular frequency of the incident
radiation. The optical parameters of the bare substrate have been
measured in a separate experiment. At 5 cm$^{-1}$ the absorptive
index of the substrate, $k_{sub}$, is below 0.01 at any
temperature. The refractive index, $n_{sub}$, is equal to
$4.06\pm0.01$ at room temperature and $4.03\pm0.01$ for $T
\rightarrow 0$ K.

\begin{figure}[]
\centering
\includegraphics[width=\columnwidth,clip]{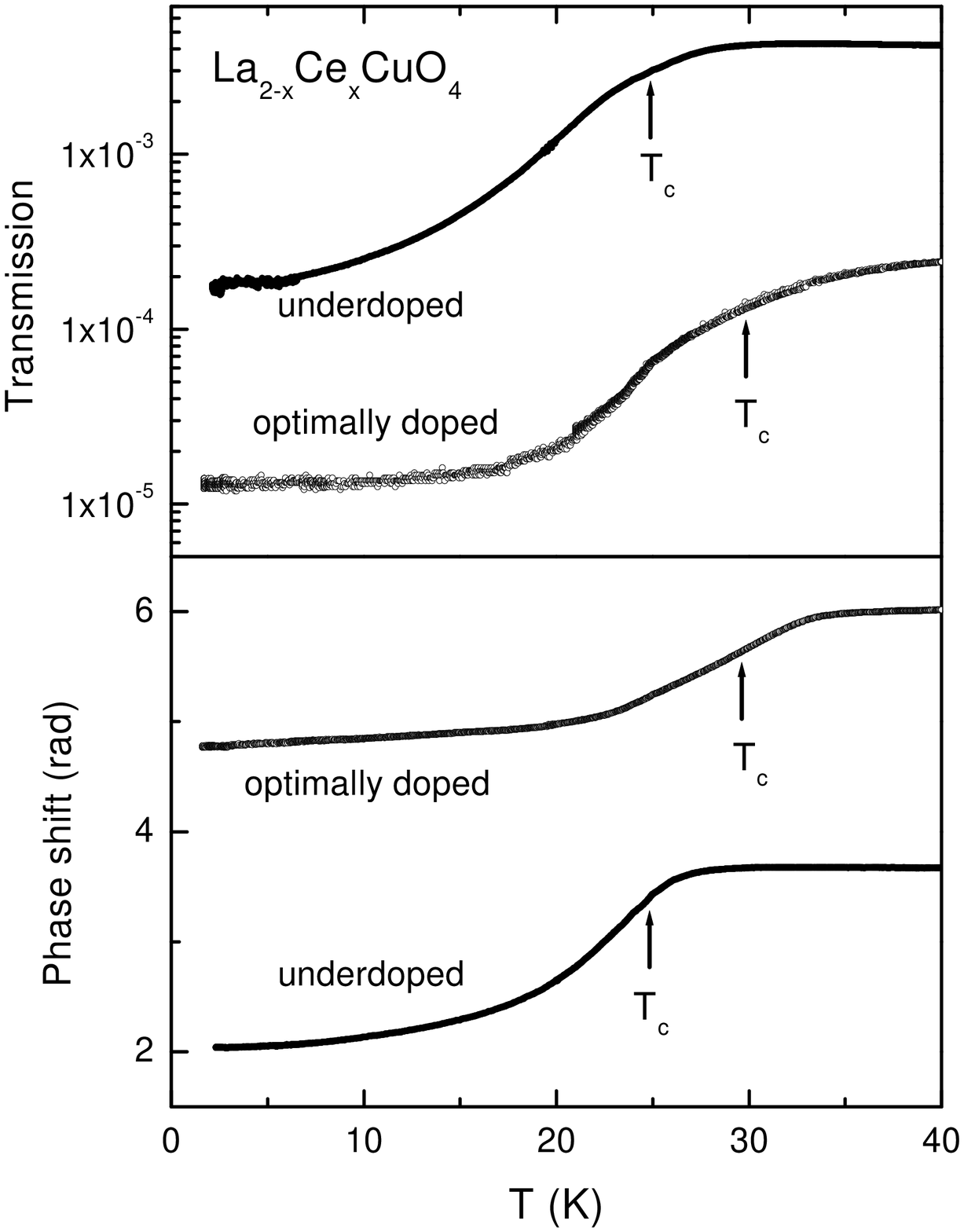}
\caption{Temperature dependence of the millimeter-wave (5
cm$^{-1}$) intensity transmission (top panel) and of the
corresponding phase shift (bottom panel) for optimally doped ($x$
= 0.106, $T_{c}$ = 30 K) and underdoped ($x$ = 0.081, $T_{c}$ = 25
K) La$_{2-x}$Ce$_{x}$CuO$_{4}$ films on SrLaAlO$_{4}$ substrates.}
\label{Tr_Pt}
\end{figure}

The as-measured transmission and phase shift data are shown in
Fig. \ref{Tr_Pt}. Since the SrLaAlO$_{4}$ substrates have very low
losses in the millimeter-to-submillimeter frequency range, the
transmission level of the substrate-film "sandwich" is determined
predominantly by the film properties. In the normally-conducting
state (above $T_{c}$) the transmission is mainly controlled by the
real part of the complex conductivity, while in the
superconducting state both components of the complex conductivity
contribute to the transmission, the imaginary part dominating at
the lowest temperatures. The overall difference in the
transmission levels between the two samples studied is due to
difference of carrier concentrations in optimally doped and
underdoped La$_{2-x}$Ce$_{x}$CuO$_{4}$.

The main contribution to the absolute values of the phase shift
comes from the substrate. The major term of the phase shift is
proportional to the substrate thickness multiplied by its
refractive index $n_{sub}$ and by the frequency of probing
radiation. Since the thickness of the substrates were different,
the phase level for the sample with optimally doped film is higher
than for the sample with the underdoped film. The film
contribution to the phase shift comes from both, the real and the
imaginary part of the complex conductivity. Below $T_{c}$ due to
the Meissner effect, $\sigma_{2}$ increases significantly and
diverges at $\omega \rightarrow 0$, that leads to the negative
contribution to the phase shift.

Let us recall that while processing the penetration depth data we
used the complete Fresnel formulas, which automatically involve
all contributions from the film complex conductivity to the
transmission and phase shift, and take multiple reflections within
the substrate into account.

\begin{figure}[]
\centering
\includegraphics[width=\columnwidth,clip]{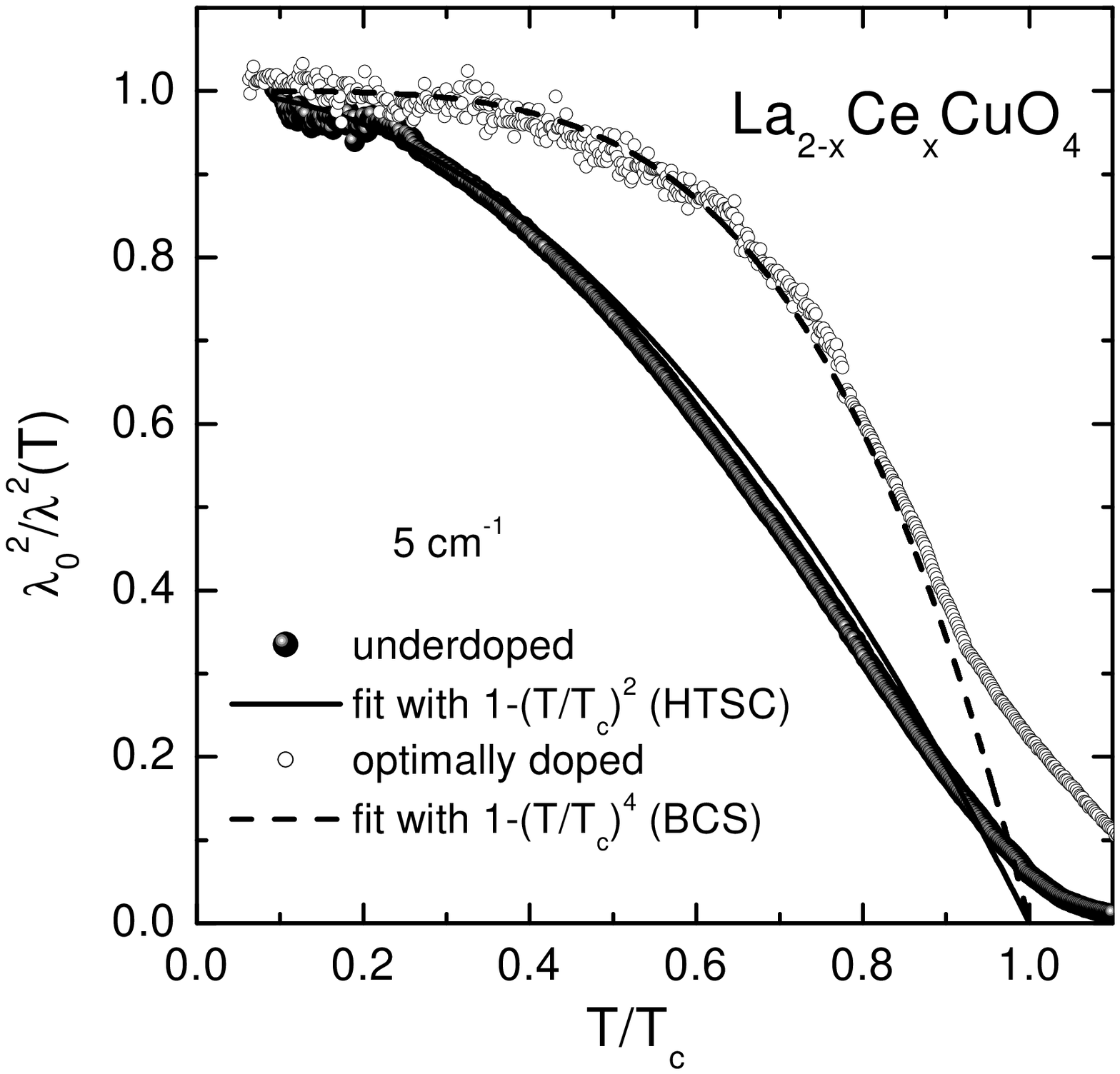}
\caption{Temperature dependence of the penetration depth in
optimally doped ($x$ = 0.106, $T_{c}$ = 30 K) and underdoped ($x$
= 0.081, $T_{c}$ = 25 K) La$_{2-x}$Ce$_{x}$CuO$_{4}$ at 5
cm$^{-1}$. Points - experimental data, lines indicate fits using
$1-(T/T_{c})^{2}$ (solid) and $1-(T/T_{c})^{4}$ (dashed).}
\label{lambda1}
\end{figure}

Fig. \ref{lambda1} shows the penetration depth of the optimally
doped and the underdoped La$_{2-x}$Ce$_{x}$CuO$_{4}$ samples
measured at 5 cm$^{-1}$. For better comparison with model
calculations, the penetration depth data are plotted as
$\lambda^{2}(0)/\lambda^{2}(T)$ versus $T/T_{c}$, where
$\lambda(0)$ is the zero-temperature limit of the penetration
depth. A substantially different behavior of $\lambda(T)$ for
samples with different doping is clearly seen at all temperatures
below $T_{c}$. In the underdoped regime the experimental curve can
be well fitted by a $[1-(T/T_{c})^{2}]$ behavior, typical for
high-temperature superconductors.\cite{t2} The results for the
optimally doped sample look quite different, now a BCS-like
\cite{t4} $[1-(T/T_{c})^{4}]$ dependence (fully gapped
superconductor) fits the experimental data. We note that the
high-temperature ($T > 0.9 T_{c}$) tails in
$\lambda^{2}(0)/\lambda^{2}(T)$ are due to influence of the
normally conducting electrons: $\sigma_{2}$ is not exactly zero
above $T_{c}$ at high frequencies.

\begin{figure}[t]
\centering
\includegraphics[width=\columnwidth,clip]{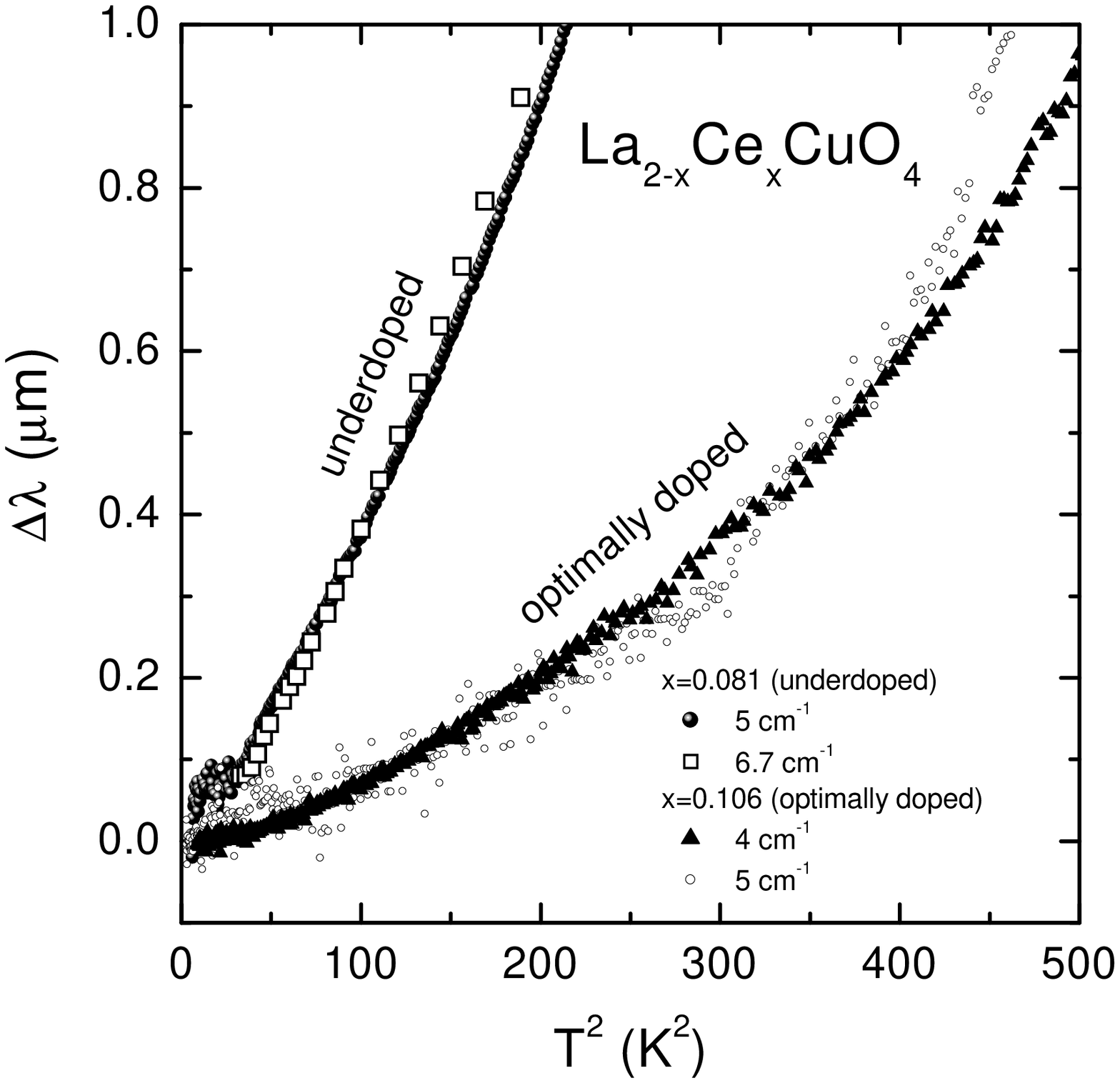}
\caption{Temperature variation of the penetration depth in
La$_{2-x}$Ce$_{x}$CuO$_{4}$ at low temperatures \textit{vs.}
$T^{2}$. The experimental data for $\nu$ = 5 cm$^{-1}$ are
replotted from Fig. \ref{lambda1}, two additional data sets at 4
and 6.7 cm$^{-1}$ are added for the optimally doped and the
underdoped samples, respectively.} \label{lambda2}
\end{figure}

The low-temperature variation of $\lambda$ is shown in Fig.
\ref{lambda2} as function of $T^{2}$. Measurements at two
additional frequencies (4 and 6.7 cm$^{-1}$) are also presented in
this figure. In this representation the data of the underdoped
sample follow rather well a straight line, while the data for the
optimally doped sample demonstrate a much more gradual (although
not really exponential) temperature dependence. No measurable
frequency dependence of our findings is detected.

The absolute values of the penetration depth at the lowest
temperatures are found to be equal to 0.22 $\pm$ 0.02 $\mu$m in
the optimally doped, and 0.38 $\pm$ 0.03 $\mu$m in the underdoped
samples. Both values are in good agreement with data by Skinta et
al. \cite{skinta} Minor differences can be explained by slightly
different doping levels of the samples marked as "underdoped" and
"optimally doped" in Ref. \onlinecite{skinta} and in the present
study.

The quadratic temperature dependence of the penetration depth in
the underdoped sample provides strong experimental evidence of
nodes in the gap function \cite{annett} and, likely, the
$d_{x^{2}-y^{2}}$-wave paring symmetry. The results in the
optimally doped sample indicate a more isotropic gap function:
$d+is$ or $s$-wave symmetry \cite{annett} seem to be good
candidates for the gap symmetry in optimally doped
La$_{2-x}$Ce$_{x}$CuO$_{4}$.

Since penetration depth measurements are not phase sensitive, it
is impossible to decide from these measurements alone, whether a
transition of the gap symmetry exists or not. However, our results
provide experimental evidence that the superconducting gap in the
underdoped regime is significantly more anisotropic than in the
optimally doped regime. As an alternative to a transition between
two superconducting states characterized by different gap
symmetries, one can imagine a gradual change in the gap anisotropy
with $d+is$, or even with just anisotropic $s$-wave, symmetry.

It is important to note that the influence of the quasiparticle
scattering on the high-frequency (millimeter-wave) penetration
depth data has to be clarified before making final conclusions on
gap symmetry or gap anisotropy transition. Strictly speaking, only
the zero-frequency penetration depth is associated with the
superconducting condensate, while the penetration depth at a
finite frequency is partly influenced by the normal-conducting
carriers.

A way to correctly and model-independently exclude the
normal-conducting contribution to the high-frequency $\lambda$ has
been recently proposed by Dordevic {\it et al.} \cite{dordevic} It
requires measurements of the complex conductivity over a wide
frequency range. Such measurements have been undertaken by us at
$T = 5$ K using the backward-wave-oscillator technique for
frequencies from 4 to 40 cm$^{-1}$, and the standard infrared
reflectivity measurements (40 - 4000 cm$^{-1}$) with a
Fourier-transform spectrometer. \cite{pimenov}

Applying the analysis given in Ref. \onlinecite{dordevic}, we have
found the corrected zero-temperature values of the penetration
depth: 0.23 $\mu$m for the optimally doped and 0.4 $\mu$m for the
underdoped samples. The minor difference between the as-measured
and the corrected penetration depth data is due to very low
residual losses in the investigated films. The details of the
recalculation procedure together with the determination of the
scattering rate in La$_{2-x}$Ce$_{x}$CuO$_{4}$ will be given
elsewhere. \cite{pronin}

In conclusion, we observed significant changes in the temperature
dependence of the millimeter-wave penetration depth in optimally
doped and underdoped La$_{2-x}$Ce$_{x}$CuO$_{4}$. $\lambda(T)$
behaves quadratic in the underdoped, but is close to exponential
in the optimally doped regime. Change in the gap anisotropy (or
symmetry) is a plausible explanation for these dramatic
differences.

We would like to thank L. Alff, J.F. Annett, D.N. Basov, G.
Duetscher, R. Gross, R. Hackl, and D. van der Marel for useful
discussions. The work was supported by BMBF via contract 13N6917/0
- EKM.


\begin{thebibliography}{99}
\bibitem{annett1} J. F. Annett, N. Goldenfeld, and A. J. Leggert,
in {\em Physical Properties of High Temperature Superconductors
V}, edited by D. M. Ginsberg (World Scientific, Singapore, 1996),
p. 375.

\bibitem{tsuei1} C. C. Tsuei and J. R. Kirtley, Rev. Mod. Phys.
\textbf{72}, 969 (2000), and references therein.

\bibitem{wu}  D. H. Wu, J. Mao, S. N. Mao, J. L. Peng, X. X. Xi, T. Venkatesan,
R. L. Greene, and S. M. Anlage, Phys. Rev. Lett. \textbf{70}, 85
(1993).

\bibitem{anlage}  S. M. Anlage, D.-H. Wu, J. Mao, S. N. Mao, X. X. Xi, T.
Venkatesan, J. L. Peng, and R. L. Greene, Phys. Rev. B
\textbf{50}, 523 (1994).

\bibitem{alff}  L. Alff, S. Kleefisch, U. Schoop, M. Zittartz, T. Kemen, T. Bauch,
A. Marx, and R. Gross, J. Eur. Phys. B \textbf{5}, 423 (1998).

\bibitem{kashiwaya}  S. Kashiwaya, T. Ito, K. Oka, S. Ueno, H. Takashima, M. Koyanagi,
Y. Tanaka, and K. Kajimura, Phys. Rev. B \textbf{57}, 8680 (1998).

\bibitem{tsuei2}  C. C. Tsuei and J. R. Kirtley, Phys. Rev. Lett. \textbf{85},
182 (2000).

\bibitem{kokales}  J. D. Kokales, P. Fournier, L. V. Mercaldo, V. V. Talanov, R. L.
Greene, and S. M. Anlage, Phys. Rev. Lett. \textbf{85}, 3696
(2000).

\bibitem{prozorov}  R. Prozorov, R. W. Giannetta, P. Fournier, and R. L. Greene, Phys.
Rev. Lett. \textbf{85}, 3700 (2000).

\bibitem{armitage}  N. P. Armitage, D. H. Lu, D. L. Feng, C. Kim, A. Damascelli, K. M.
Shen, F. Ronning, Z.-X. Shen, Y. Onose, Y. Taguchi, and Y. Tokura,
Phys. Rev. Lett. \textbf{86}, 1126 (2001).

\bibitem{sato}  T. Sato, T. Kamiyama, T. Takahashi, K. Kurahashi, and K. Yamada,
Science \textbf{291}, 1517 (2001).

\bibitem{deutscher}  Y. Dagan and G. Deutscher, Phys. Rev. Lett. \textbf{87},
177004 (2001).

\bibitem{sharoni}   A. Sharoni, O. Millo, A. Kohen, Y. Dagan,
R. Beck, G. Deutscher, and G. Koren, Phys. Rev. B \textbf{65},
134526 (2002).

\bibitem{yeh}  N.-C. Yeh, C.-T. Chen, G. Hammerl, J. Mannhart,
A. Schmehl, C. W. Schneider, R. R. Schulz, S. Tajima, K. Yoshida,
D. Garrigus, and M. Strasik, Phys. Rev. Lett. \textbf{87}, 087003
(2001).

\bibitem{skinta}  J. A. Skinta, M.-S. Kim, T. R. Lemberger, T. Greibe, and M. Naito,
Phys. Rev. Lett. \textbf{88}, 207005 (2002).

\bibitem{biswas}  A. Biswas, P. Fournier, M. M. Qazilbash,
V. N. Smolyaninova, H. Balci, and R. L. Greene, Phys. Rev. Lett.
\textbf{88}, 207004 (2002).

\bibitem{laughlin} R. B. Laughlin, Phys. Rev. Lett. \textbf{80},
5188 (1998).

\bibitem{vojta} M. Vojta, Y. Zhang, and S. Sachdev,
Phys. Rev. Lett. \textbf{85}, 4940 (2000).

\bibitem{khveshchenko} D. V. Khveshchenko and J. Paaske,
Phys. Rev. Lett. \textbf{86}, 4672 (2001).

\bibitem{kozlov}  G. V. Kozlov and A. A. Volkov, in {\it Millimeter and
Submillimeter Wave Spectroscopy of Solids}, edited by G.
Gr\"{u}ner (Springer, Berlin, 1998), p. 51.

\bibitem{naito1} M. Naito and M. Hepp, Physica C \textbf{357-360}, 333 (2001).

\bibitem{naito2} M. Naito and M. Hepp, Jpn. J. Appl. Phys. \textbf{39}, L485 (2000).

\bibitem{yamamoto}  H. Yamamoto, M. Naito, and H. Sato, Phys. Rev.
B \textbf{56}, 2852 (1997).

\bibitem{t2} D. A. Bonn and W. N. Hardy in {\em Physical Properties
of High Temperature Superconductors V}, edited by D. M. Ginsberg
(World Scientific, Singapore, 1996), p. 7.

\bibitem{t4}  M. Tinkham, {\it Introduction to Supercondictivity} (McGraw-Hill, New York,
1975).

\bibitem{annett} J. F. Annett, N. D. Goldenfeld, and S. R. Renn, in {\it Physical
Properties of High-Temperature Superconductors II}, edited by D.
M. Ginsberg (World Scientific, Singapore, 1990), p. 571; J.
Annett, N. Goldenfeld, and S. R. Renn, Phys. Rev. B \textbf{43},
2778 (1991).

\bibitem{dordevic} S. V. Dordevic, E. J. Singley, D. N. Basov,
S. Komiya, Y. Ando, E. Bucher, C. C. Homes, and M. Strongin, Phys.
Rev. B \textbf{65}, 134511 (2002).

\bibitem{pimenov} For the underdoped composition these
measurements have been reported in: A. Pimenov, A. V. Pronin, A.
Loidl, A. Tsukada, and M. Naito, cond-mat/0212400.

\bibitem{pronin} A. V. Pronin, A. Pimenov, A. Loidl, A. Tsukada,
and M. Naito, unpublished.

\end{thebibliography}
\end{document}